\documentstyle[epsf,twocolumn,aps]{revtex}
\begin{document}
\twocolumn[\hsize\textwidth\columnwidth\hsize\csname @twocolumnfalse\endcsname

\tightenlines
%\draft

\title{Electronic Structure of Cu$_{1-x}$Ni$_x$Rh$_2$S$_4$ and CuRh$_2$Se$_4$: Band Structure 
Calculations, X-ray Photoemission and Fluorescence Measurements}

% Repeat the \author\address pair as needed

\author{G. L. W. Hart, W. E. Pickett}
\address{Department of Physics, University of California, Davis, 95616-8677, USA}

\author{ E. Z. Kurmaev}
\address{Institute of Metal Physics, Russian Academy of Sciences-Ural Division,
620219 Yekaterinburg GSP-170, Russia}

\author{D. Hartmann, M. Neumann}
\address{University of Osnabr\"uck, Department of Physics, Osnabr\"uck D-49069, Germany}

\author{A. Moewes} 
\address{CAMD/LSU, Baton Rouge, LA 70803, USA}

\author{D. L. Ederer}
\address{Department of Physics, Tulane University, New Orleans, LA 70118, USA}

\author{R. Endoh, K. Taniguchi, and S. Nagata}
\address{Department of Materials Science and Engineering, Muroran Institute
of Technology, Muroran 050-8585, Japan}

\date{\today}
\maketitle

\begin{abstract}
The electronic structure of spinel-type Cu$_{1-x}$Ni$_x$Rh$_2$S$_4$ (x = 0.0, 0.1, 0.3, 
0.5, 1.0) and CuRh$_2$Se$_4$ compounds has been studied by means of X-
ray photoelectron and fluorescent spectroscopy. Cu L$_3$, Ni L$_3$, S L$_{2,3}$ 
and Se M$_{2,3}$ X-ray emission spectra (XES) were measured near 
thresholds at Beamline 8.0 of the Lawrence Berkeley Laboratory's 
Advanced Light Source. XES measurements of the constituent atoms 
of these compounds, reduced to the same binding energy scale, are
found to be in excellent agreement with XPS valence bands. 
The calculated XES spectra which include dipole matrix elements 
show that the partial density of states reproduce experimental spectra 
quite well. States near the Fermi level (E$_F$) have strong Rh d and
S(Se) p character in all compounds. In NiRh$_2$S$_4$ the Ni 3d states contribute
strongly at E$_F$, whereas in both Cu compounds the Cu 3d bands are only $\sim$1 eV wide and
centered $\sim$2.5 eV below E$_F$, leaving very little 3d character at E$_F$.
 The density of states at the Fermi level is less in 
NiRh$_2$S$_4$ than in CuRh$_2$S$_4$. This difference may contribute to the 
observed decrease, as a function 
of Ni concentration, in the 
superconducting transition temperature in Cu$_{1-x}$Ni$_x$Rh$_2$S$_4$. The density of states of the ordered alloy Cu$_{0.5}$Ni$_{0.5}$Rh$_2$S$_4$
shows behavior that is more ``split-band''-like than ``rigid band''-like.
\end{abstract}

%This turns two column back on
]

\section{Introduction}
\label{intro}
Spinel compounds exhibit an extensive variety of interesting 
physical properties and have potential technological applications. There are a variety of
3d ion-based oxide spinels, while the S and Se counterparts usually contain 4d or 5d atoms.
Several 
of the compounds are superconductors (LiTi$_2$O$_4$, CuRh$_2$S$_4$, CuRh$_2$Se$_4$, etc.), 
there are unusual magnetic insulators (e.g.\ LiMn$_2$O$_4$ and Fe$_3$O$_4$), and 
recently, the first d-electron-based heavy fermion metal has been 
discovered (LiV$_2$O$_4$).\cite{kondo} The suprisingly high value of the 
superconducting critical temperature (11 K) in LiTi$_2$O$_4$ has never been understood.\cite{johnston}
 Another spinel compound, CuIr$_2$S$_4$, is neither 
magnetic nor superconducting but displays a rather unusual metal-insulator 
transition that is not yet understood.\cite{nagata} The ternary sulfo- and 
selenospinels CuRh$_2$S$_4$ and CuRh$_2$Se$_4$ have 
been found to be superconducting at T$_c$ = 4.70 and 3.48 K, respectively.\cite{hagino}  
They have the typical spinel structure [$Fd\overline{3}m$] where Cu ions occupy the $A$ 
tetrahedral sites and Rh ions occupy the $B$ octahedral sites. 

This wide range of phenomena in the spinel-structure oxide compounds raises
very general questions about the electronic structure of the sulfides and the 
selenides: are there indications
of strong correlations effects, or can their properties be accounted for as Fermi
liquids described by conventional band theory?
Different models 
for the valence of Cu in these compounds have been discussed,\cite{goodenough,lotgering}  but 
according to recent photoemission measurements given for CuV$_2$S$_4$,\cite{lu}  
CuIr$_2$S$_4$, CuIr$_2$Se$_4$\cite{matsuno} and Cu$_{0.5}$Fe$_{0.5}$Cr$_2$S$_4$,\cite{postnikov}
Cu is best characterized as 
monovalent in spinel compounds. Therefore, one expects that the Rh ion 
will have a formal mixed valence of +3.5 in CuRh$_2$S$_4$ and CuRh$_2$Se$_4$, and 
indeed both are good metals. However, very little of the typical 
temperature-dependent behavior of ``mixed valence compounds'' is seen in these Rh-based spinels.

The electrical and magnetic properties of Cu$_{1-x}$Ni$_x$Rh$_2$S$_4$ have been 
presented by Matsumoto et al.\cite{matsumoto}  The superconducting transition temperature 
decreases (4.70 K $\rightarrow$ 3.7 K $\rightarrow$ 2.8 K $\rightarrow$ $<$ 2.0 K) as Cu is replaced by Ni (x 
= 0.00, 0.02, 0.05, and 0.10), but the reason for this behavior is 
unexplained. Hagino et al.\cite{hagino} have presented extensive data on CuRh$_2$S$_4$ and 
CuRh$_2$Se$_4$ (resistivity, susceptibility, magnetization, specific heat, NMR), 
but their differences do not yet have any microscopic interpretation. Only 
for CuRh$_2$S$_4$ have general (full potential, all electron) band structure 
calculations been reported.\cite{oda}

In this paper, we present X-ray spectroscopic studies of the valence 
band electronic structure of these materials. To provide a clear 
interpretation of these data, we also report first-principles band structure 
calculations (Linear-Augmented-Plane-Wave method [LAPW]) for 
CuRh$_2$S$_4$, CuRh$_2$Se$_4$, NiRh$_2$S$_4$ and Cu$_{0.5}$Ni$_{0.5}$Rh$_2$S$_4$ that enable us to address 
the properties of these spinels. Total and partial densities of states (DOS), 
plasma energies and transport-related quantities are calculated as well as X-ray
 emission spectra. The total and partial DOS and calculated X-ray 
emission spectra are found to compare favorably with the measured X-ray 
photoelectron spectra (XPS) and X-ray emission spectra (XES) (which 
probe total and partial DOS, respectively). All spectral measurements are 
performed using the same samples which were used to study the electrical 
and magnetic properties of Cu$_{1-x}$Ni$_x$Rh$_2$S$_4$ in Ref.\ \protect\onlinecite{matsumoto}.

\section{Experimental Details}
\label{exp}
Mixtures of high-purity fine powders of Cu, Ni, Rh, S and Se with 
nominal stoichiometry were heated in sealed quartz tubes at $850^{\circ}$ C for a 
period of 10 days. Subsequently, the specimens were reground and sintered 
in pressed parallelepiped form at $850^{\circ}$ C for 48 hours. X-ray diffraction 
data confirms the spinel phase in these powder specimens. The lattice 
constants of Cu$_{1-x}$Ni$_x$Rh$_2$S$_4$ are 9.79, 9.79 and 9.71 \AA\ for x = 0.0, 0.1 and 
1.0, respectively, and 10.27 \AA\ for CuRh$_2$Se$_4$.

The XPS measurements were performed with an ESCA 
spectrometer from Physical Electronics (PHI 5600 ci, with 
monochromatized Al-K$_\alpha$ radiation of a 0.3 eV FWHM). The energy 
resolution of the analyzer was 1.5$\%$ of the pass energy. The estimated 
energy resolution was less than 0.35 eV for the XPS measurements on the 
copper and nickel sulfides. The pressure in the vacuum chamber during the 
measurements was below $5 \times 10^{-9}$ mbar. Prior to XPS measurements the 
samples were cleaved in ultra high vacuum. All the investigations have 
been performed at room temperature on the freshly cleaved surface. The 
XPS spectra were calibrated using a Au-foil to obtain photoelectrons from 
the Au-4f$_{7/2}$ subshell. The binding energy for Au-4f$_{7/2}$ electrons is 84.0 eV.

X-ray fluorescence spectra were measured at Beamline 8.0 of the 
Advanced Light Source (ALS) at Lawrence Berkeley Laboratory. The 
undulator beam line is equipped with a spherical grating monochromator,\cite{jia}  
and an experimental resolving power of $E/\Delta E = 300$ was used. The 
fluorescence end station consists of a Rowland circle grating spectrometer. 
The Ni L$_3$ and Cu L$_3$ XES were measured with an experimental resolution 
of approximately 0.5--0.6 eV and S L$_{2,3}$ and Se M$_{2,3}$ with resolution of 0.3--0.4 eV.
The incident angle of the $p$-polarized monochromatic beam on the 
sample was about $15^{\circ}$. The Cu L$_3$ and Ni L$_3$ XES were measured just above 
L$_3$ threshold but below the L$_2$ threshold which prevented overlap of the 
metal L$_3$ and metal L$_2$ spectra.

\section{Method of Calculation}
\label{calcs}
The band structure calculations were done 
with the full potential LAPW code WEIN97.\cite{blaha}  The 
sphere radii were chosen as 2.1 a.u., 2.2 a.u., 
and 2.0 a.u.\ for Cu/Ni, Rh, and S/Se, 
respectively. The plane wave cutoff was K$_{max}$ = 3.25 
a.u., resulting in slightly more than 1400 basis 
functions per primitive cell ($\sim$100 basis 
functions/atom). The local density approximation 
(LDA) exchange-correlation potential of Perdew and 
Wang\cite{pw} was used  except for the DOS calculations 
shown in Fig.\ 7 where the gradient correction to 
the LDA exchange-correlation potential of Perdew, 
Burke, and Ernzerhof\cite{pbe}  was used. A mesh of 47 k-points
in the irreducible zone (Bl\"ochl's modified 
tetrahedron method\cite{blochl}) was used in achieving self-
consistency.

The XES spectra were calculated using Fermi's 
golden rule and the matrix elements between the 
core and valence states (following the formalism 
of Neckel\cite{neckel}). The calculated spectra include 
broadening for the spectrometer and core and 
valence lifetimes. The DOS calculations used
47 k-points (again, Bl\"ochl's modified tetrahedron 
method was used). The experimental lattice 
constants (listed in the previous section) were 
used in the calculations and the values used for 
the internal parameter $u$ were taken to be 0.385 
for all three stoichiometric compounds (CuRh$_2$Se$_4$, 
CuRh$_2$S$_4$, NiRh$_2$S$_4$) as well as for Cu$_{0.5}$Ni$_{0.5}$Rh$_2$S$_4$. 
Experimental data for the internal parameter was 
not available, so the values were taken to be 
0.385 (rather than the ``ideal'' position of 3/8) by 
analogy to the related CuIr$_2$S$_4$ and CuIr$_2$Se$_4$ spinel 
compounds for which the $u$ parameter has been 
measured.

\section{Discussion of Spectroscopic Data}
\subsection{CuRh$_2$S$_4$ and NiRh$_2$S$_4$}

The calculated total and partial DOS of CuRh$_2$S$_4$ and NiRh$_2$S$_4$,
shown in Figs.\ \ref{cus} and \ref{nis}, reveal many 
common 
features. The valence bands extend from E$_F$ (taken as the zero of energy) to 
approximately -7 eV and the Fermi level lies near the top of a Rh d-chalcogen
p complex of bands that lie below a gap centered 0.5--1.0 eV 
above the Fermi level. The gap between the valence band and conduction 
band is found to be about of 0.5--0.7 eV wide. The sulfur states in CuRh$_2$S$_4$ 
and NiRh$_2$S$_4$ show similar DOS, S 3s atomic-like states in the region -12.7 $\sim$
 -14.7 eV and band-like S 3p states which are mixed with Rh 4d and 
Cu/Ni 3d-states in a wide energy region. Cu/Ni 3d-states are found to be 
much narrower than Rh 4d states which are less localized and form two 
peaks in the DOS near the bottom and the top of the valence band. Our 
results for CuRh$_2$S$_4$ are similar to those of Ref.\ 8 except for the distribution 
of Cu 3d DOS.\cite{note}  As seen in Fig.\ \ref{cus}, Cu 3d states lie within 
the region of S 
3p states but are weakly hybridized, forming a 1 eV wide peak centered 
around -2.5 eV. The S d character is quite small and probably reflects tails 
of the neighboring atoms more than atomic 3d character.

The total DOS at the Fermi level [N(E$_F$)] increases from NiRh$_2$S$_4$ 
(8.18 states/eV/cell) to CuRh$_2$S$_4$ (9.89 states/eV/cell) which has the same trend as
electronic specific heat coefficients measured in Refs.\ \onlinecite{hagino} and
 \onlinecite{nagata2}.
For the intermediate compound 
Cu$_{0.5}$Ni$_{0.5}$Rh$_2$S$_4$, N(E$_F$) is 8.43 states/eV/cell, much nearer that of
NiRh$_2$S$_4$. In CuRh$_2$S$_4$ the 
Cu 3d partial DOS is very small at the Fermi level whereas Rh 4d and S 3p 
partial DOS are the main contribution to the total. Consequently,
the Cooper pairs in the superconducting state of CuRh$_2$S$_4$ are 
formed mainly by the electrons in the hybridized bands derived from Rh 4d 
and S 3p states. 

In NiRh$_2$S$_4$ the situation is quite different. Ni 3d states are broader 
and at lower binding energy than the Cu 3d states of CuRh$_2$S$_4$,
 and hybridization with S p leads to Ni 
3d character over a 3 eV wide region that extends above the Fermi level. 
The result is that the main contribution to the DOS at the Fermi level is 
from Ni 3d states, unlike in CuRh$_2$S$_4$ where the Cu 3d contribution at E$_F$ is very minor.

The experimental Cu L$_3$ ($3d4s\rightarrow 2p$ transition), Ni L$_3$ ($3d4s\rightarrow 2p$ 
transition) and S L$_{2,3}$ ($3s3d\rightarrow 2p$ transition) XES
probe Cu 3d4s, Ni 3d4s and S 3s3d partial DOS in the valence band and, in 
the first approximation, can be directly compared with calculated band 
structures. The comparison of the calculated and measured partial DOS are 
shown in Figs.\ \ref{cusxes} and \ref{nisxes}, where Cu L$_3$, Ni L$_3$ and S L$_{2,3}$ XES are converted to 
the binding energy scale using our XPS measurements of the corresponding 
core levels [E$_{b.e.}$(Cu 2p) = 932.39 eV, E$_{b.e.}$(Ni 2p) = 852.98 eV and E$_{b.e.}$(S 
2p) = 161.57 eV]. We see that the measured Cu L$_3$, Ni L$_3$ and S L$_{2,3}$ XES 
peaks are very close to Cu 3d, Ni 3d and S 3s partial DOS in CuRh$_2$S$_4$ (Fig.\ \ref{cusxes})
and NiRh$_2$S$_4$ (Fig.\ \ref{nisxes}). In each case, the peaks in the calculated DOS lie at 
somewhat lower binding energy:~1 eV for S 3s and Cu 3p, but only a few 
tenths of eV for Ni 3d. The difference reflects a self-energy correction that 
lies beyond our band theoretical methods. In addition, we calculated the 
emission intensities of Cu/Ni L$_3$, Rh N3 ($4d\rightarrow 4p$ transition)\cite{foot1} and S L$_{2,3}$ 
XES in both compounds as described in section \ref{calcs}. The calculated spectra 
are presented in the same figures (Figs.\ 3 and 4) and show close correspondence
with experimental spectra as well as with the corresponding 
partial DOS. From the close agreement, we conclude that the influence of 
core holes in the measured XES spectra is minor and experimental spectra 
can be understood directly from the calculated spectra and partial DOS.

\subsection{Cu$_{1-x}$Ni$_x$Rh$_2$S$_4$ (x = 0, 0.1, 0.3, 0.5, 1.0)}
We measured XPS valence band (VB) spectra for the Cu$_{1-x}$Ni$_x$Rh$_2$S$_4$ (x = 
0, 0.1, 0.3, 0.5, 1.0) system (see Fig.\ \ref{cunixps}) and found a four peak structure: ($a$, 
$c$, $d$, $e$) for CuRh$_2$S$_4$ and ($a$, $b$, $d$, $e$) for NiRh$_2$S$_4$, each of which is very close to 
the corresponding calculated total DOS (Figs.\ \ref{cus} and \ref{nis}). Based on our calculations, we can 
conclude that the $a$ peak at 1 eV binding energy is formed by Rh 4d-S 3p 
states for CuRh$_2$S$_4$ and Ni 3d-Rh 4d--S 3p states for NiRh$_2$S$_4$. The next 
peak ($b$ for NiRh$_2$S$_4$ at 2 eV binding energy and $c$ for CuRh$_2$S$_4$ at 3 eV 
binding energy) can be attributed mainly to Ni (respectively Cu) 3d states. The $d$ peak 
(5.5 eV) relates to Rh 4d-S 3p states and the $e$ peak is associated with 
atomic-like S 3s states. In the solid solution Cu$_{1-x}$Ni$_x$Rh$_2$S$_4$ the positions of 
the peaks do not change as the concentration varies, but only the ratio 
of intensities of $b$ (Ni 3d) and $c$ (Cu 3d) peaks vary according to the Cu/Ni 
concentration. 

This behavior suggests that the electronic structure of the solid 
solution Cu$_{1-x}$Ni$_x$Rh$_2$S$_4$ can be deduced by analyzing the endpoints (x = 0.0 
and 1.0), CuRh$_2$S$_4$ and NiRh$_2$S$_4$. This conclusion results not from a rigid 
band picture (which does not hold) but from the opposite ``split-band'' 
behavior in which both Cu and Ni retain their own DOS peaks which then 
vary in strength roughly as the concentration. In Fig.\ \ref{cunixes} we have compared 
XPS VB measurements with Cu L$_3$, Ni L$_3$ and S L$_{2,3}$ XES spectra for 
Cu$_{0.5}$Ni$_{0.5}$Rh$_2$S$_4$.\cite{foot2} We see that positions of the peaks in the Ni L$_3$, Cu 
L$_3$ and S L$_{2,3}$ XES spectra correspond exactly to peaks $b$, $c$ and $e$ of the 
XPS VB measurements, which is consistent with our interpretation of the 
XPS data as indicating a solid solution of Cu$_{1-x}$Ni$_x$Rh$_2$S$_4$ if the split-band 
behavior holds. 

In Fig.\ \ref{finedos} we have compared the calculated total DOS of CuRh$_2$S$_4$, 
NiRh$_2$S$_4$, and Cu$_{0.5}$Ni$_{0.5}$Rh$_2$S$_4$. With respect to the top of the highest 
occupied bands, the Fermi energy is highest in the bands of CuRh$_2$S$_4$ to 
accommodate the two additional electrons from the Cu atoms. The 
behavior of the DOS for the three systems shown are quite different, 
particularly for Cu and Ni ions, in an energy range between the Fermi 
levels for NiRh$_2$S$_4$ and for CuRh$_2$S$_4$, invalidating a rigid-band interpretation 
of the differences and similarities in these compounds. This is not 
surprising given the different character of the Ni- and Cu-derived states in 
this energy region. As mentioned above, whereas states at the Fermi level 
in NiRh$_2$S$_4$ have a strong Ni 3d character, Cu 3d states lie entirely
below the Fermi level in CuRh$_2$S$_4$. The character of 
states at the Fermi level in CuRh$_2$S$_4$ are primarily Rh d-like states 
hybridized with S 3p states. 

According to Ref.\ \onlinecite{matsumoto}, the superconducting transition temperature of 
Cu$_{1-x}$Ni$_x$Rh$_2$S$_4$ decreases with increasing Ni concentration from 4.7 K (x = 
0.0) to 3.7 K (x = 0.02) and then to 2.8 K (x = 0.05). While we attribute this 
to a general decrease in DOS at the Fermi level as the Ni concentration is 
increased (see Sec. \ref{other}), this trend does not require a simple rigid-band interpretation. In 
the alloy, the DOS within a few tenths of an eV of E$_F$ probably cannot be 
described by either the rigid band or split-band models.

\subsection{CuRh$_2$Se$_4$}
Figure \ref{cuse} shows the calculated total and partial DOS for CuRh$_2$Se$_4$. 
While it is similar to that of CuRh$_2$S$_4$ (Fig.\ 1), we can point out two 
differences: ($i$) the Se 4p DOS is redistributed somewhat compared to 
S 3p and has a higher contribution in the vicinity of the Fermi 
level, and ($ii$) the Se d-like character is even less than that of the d-like 
character in CuRh$_2$S$_4$. The total DOS at the Fermi level is 12.05 states/eV-cell
which is higher than in CuRh$_2$S$_4$, in qualitative agreement with 
measurements of electronic specific-heat measurements.\cite{hagino}

In Fig.\ \ref{cusexes} the 
experimental Cu L$_3$ and Se M$_{2,3}$ ($4s\rightarrow 3p$ transition) XES measurements are 
compared to the Cu 3d and Se 4s partial DOS and calculated spectra. The 
agreement of the peak positions between experiment and theory is quite close. Again we note 
that calculated XES spectra exactly follow the partial DOS, as in the case 
of CuRh$_2$S$_4$ and NiRh$_2$S$_4$ (Figs.\ \ref{cusxes} and \ref{nisxes}). The XPS valence band data is 
compared with the Cu L$_3$ and Se M$_{2,3}$ XES spectra of Fig.\ \ref{cusexps}.
The location of Cu 3d-Se 4s-derived bands is reproduced well 
(comparable to that in the sulfide) by the calculations. There are some 
differences in ratio of the XPS peaks for CuRh$_2$Se$_4$ and CuRh$_2$S$_4$: the 
relative intensity of Cu 3d peak located of around 2.5 eV is less in 
CuRh$_2$Se$_4$ than in CuRh$_2$S$_4$. This may be due to the 2.5 times larger photo-ionization
 cross-section of Se 4p states as compared to that of S 3p states.\cite{yeh} 

\section{Other Data}
\label{other}
	In a metal the Drude plasma energy tensor $\hbar\Omega_{p,ij}$ contains a good 
deal of information about low temperature transport and low frequency 
optical properties. $\Omega_{p,ij}$ is given by

\begin{eqnarray}
\Omega _{p,ij}^2 & = & 4\pi e^2{1 \over V}\sum\limits_k {v_{k,i}v_{k,j}\delta
 (\varepsilon _k-\varepsilon _F)}\nonumber \\
 & = & 4\pi e^2\left\langle {v_iv_j} \right\rangle N(\varepsilon_F)
\label{omega}
\end{eqnarray}
where $v_{k,i}$ is the i-th cartesian coordinate of the electron velocity, $V$ is the normalization 
volume and $\langle\cdots\rangle$ indicates a Fermi surface average. The optical conductivity (specializing 
now to cubic metals) contains a $\delta$-function contribution at zero frequency proportional 
to $\Omega_p^2$ (which is broadened by scattering processes), and the static 
conductivity in Bloch-Boltzmann theory\cite{ziman} becomes

\begin{equation}
\rho (T)=\rho _0+{{4\pi } \over {\Omega _p^2\tau }}
\label{rho}
\end{equation}
($\rho_0$ is the residual resistivity at T = 0) as long as the mean free path $l = v_F\tau$ is large 
enough that scattering processes are independent. When phonon scattering dominates, 
which is usually the case above 25$\%$ of the Debye temperature, the relaxation time $\tau$
becomes approximately\cite{allen}

\begin{equation}
{\hbar  \over {\tau _{ep}}}=2\pi \lambda _{tr}k_BT
\label{tau}
\end{equation}
where $\lambda_{tr}$ is a ``transport'' electron-phonon (EP) coupling strength that is usually close to 
the EP coupling constant $\lambda$ that governs superconducting properties. Then in the high T 
regime we obtain the estimate

\begin{equation}
\lambda \approx \lambda _{tr}\approx {{\hbar \Omega _p^2} \over {8\pi ^2k_B}}{{d\rho } \over {dT}}
\label{lambda}
\end{equation}. 

Hagino et al.\onlinecite{hagino} have presented resistivity data on sintered samples of CuRh$_2$S$_4$ and 
CuRh$_2$Se$_4$. Although both are clearly metallic (d$\rho$/dT $>$ 0), the magnitudes of $\rho$ differ by 
a factor of 20 over most of the range 50 K $\leq$ T $\leq$ 300 K. CuRh$_2$Se$_4$ has $\rho_0 = 2~\mu\Omega cm$, 
indicating excellent metallic behavior in spite of the intergrain scattering that is present in 
the  sintered samples. The CuRh$_2$S$_4$ sample had $\rho_0 = 500~\mu\Omega cm$ (perhaps from intergrain 
scattering connected to differences in surface chemistry of the sulfide and the selenide) 
which makes Eq.\ (\ref{rho}) inapplicable. Moreover, both materials (especially CuRh$_2$S$_4$) show 
saturation behavior which makes the Bloch-Boltzmann analysis less definitive. However, 
we can apply this formalism to CuRh$_2$Se$_4$ to obtain an estimate, using $d\rho/dT \approx 2 \mu\Omega cm/K $
to obtain $\lambda_{tr} = 1.8$. This value is almost a factor of three larger than $\lambda = 0.64$ found by 
Hagino et al.\ to be sufficient to account for T$_c$ = 3.5 K. We expect that the magnitude of $\rho$ 
measured on the sintered sample of CuRh$_2$Se$_4$, although small, is still not representative of the bulk.

From their measurements, Hagino et al.\cite{hagino}\ inferred almost indistinguishable values 
of the linear specific heat coefficient $\gamma$, the density of states N(E$_F$), and electron-phonon 
coupling strengths $\lambda$ for CuRh$_2$S$_4$ and CuRh$_2$Se$_4$. Our calculations lead to a 20$\%$ higher 
value of N(E$_F$) in the selenide which is at odds with their values. The 1.2 K lower value of 
T$_c$ in the selenide is not very definitive, since this difference could be related to softer 
phonon frequencies. The nearly factor of two increase in the susceptibility in the selenide 
(and not in the sulfide) below 300 K remains unexplained. Data on single crystal samples 
may be necessary to resolve these discrepancies.

\section{Conclusions}

The main results of the present study of the electronic structure in Cu$_{1-x}$Ni$_x$Rh$_2$S$_4$ 
and CuR$_2$Se$_4$ can be summarized as follows. The electronic states near E$_F$ consist mainly 
Rh 4d and S(Se) 3p(4p) orbitals for CuRh$_2$S$_4$ and CuRh$_2$Se$_4$ and primarily Ni 3d with 
some Rh 4d and S 3p orbitals in NiRh$_2$S$_4$. Thus, we find that character of states at the 
Fermi level changes in a non-rigid-band way in Cu$_{1-x}$Ni$_x$Rh$_2$S$_4$, and while there is a 
general trend of a decreasing DOS at the Fermi level as a function of Ni concentration, 
we have found that the superconducting trends in Cu$_{1-x}$Ni$_x$Rh$_2$S$_4$ cannot be explained 
quantitatively by the calculated DOS of the Cu$_{1-x}$Ni$_{x}$Rh$_2$S$_4$ system.
Moreover, such an 
interpretation would be at odds with the partial DOS which 
shows the different character of states near E$_F$. The measured X-ray data suggests 
interpreting Cu$_{1-x}$Ni$_x$Rh$_2$S$_4$ as a solid state solution more in line with a ``split-band'' 
interpretation.

Calculated X-ray emission spectra are found to be in an excellent agreement with 
experimental data, with peak positions differing by only 0.3--1.0 eV. This agreement 
implies that core hole effects are negligible. In addition to total DOS,
 plasma energies have been calculated and used to offer 
additional theoretical input to interpret the differences between CuRh$_2$S$_4$ and CuRh$_2$Se$_4$. 
Unfortunately, transport data appear to be too strongly affected by intergrain scattering to 
allow a quantitative analysis.

To summarize, the very good agreement between the measured and calculated electronic spectra
indicate a lack of any strong correlation effects. The decrease in
superconducting T$_c$ with Ni concentration is likely due to a decrease in N(E$_F$).
Beyond these general conclusions, however, several questions remain. The linear specific 
heat coefficients are not accounted for quantitatively; neither are the intermediate 
temperature resistivities, but these must be measured on single crystals to obtain a good experimental picture.
Finally, the temperature dependence of the susceptibility of CuRh$_2$Se$_4$ remains unexplained.
 
\section{Acknowledgments}

This work was supported by the Russian Science Foundation for Fundamental 
Research (Projects 96-15-96598 and 98-02-04129), a NATO Linkage Grant (HTECH.LG 
971222), INTAS-RFBR (95-0565), NSF Grants (DMR-9017997, DMR-9420425, and 
DMR-9802076), and the DOE EPSCOR and Louisiana Education Quality Special Fund 
(DOE-LEQSF (1993-95-03)). Work at the Advanced Light Source at Lawrence Berkeley 
National Laboratory was supported by the U.S. Department of Energy under contract 
(DE-AC03-76SF00098).

% FIG. 1
\begin{figure}
\caption{Calculated total (top panel) and partial DOS in CuRh$_2$S$_4$. Note the hybridization 
gap that lies just above the Fermi level (taken as the zero of energy), indicating the Fermi 
level lies in a bonding region of the electronic structure.}
\label{cus}
\end{figure}

% FIG. 2
\begin{figure}
\caption{Calculated total (top panel) and partial DOS as in Fig.\ 1 but for NiRh$_2$S$_4$.}
\label{nis}
\end{figure}

% FIG. 3
\begin{figure}
\caption{Comparison of calculated XES and partial DOS with experimental spectra of 
CuRh$_2$S$_4$. Calculations used the LAPW method as described in the text.}
\label{cusxes}
\end{figure}

\begin{figure}
\caption{Comparison of calculated XES and partial DOS with experimental spectra of
NiRh$_2$S$_4$.}
\label{nisxes}
\end{figure}

\begin{figure}
\caption{XPS VB of Cu$_{1-x}$Ni$_x$Rh$_2$S$_4$ (x = 0.0, 0.1, 0.3, 0.5, 1.0).
The peaks and shoulders  $a, b, c, d, e$ are discussed in the text.}
\label{cunixps}
\end{figure}

\begin{figure}
\caption{Comparison of the valence band XPS spectrum (upper set of data) to the
Cu L$_3$, Ni L$_3$ and S L$_{2,3}$ XES in Cu$_{0.5}$Ni$_{0.5}$Rh$_2$S$_4$.
Note the close alignment of XPS and XES peaks.}
\label{cunixes}
\end{figure}

\begin{figure}
\caption{Calculated total DOS of NiRh$_2$S$_4$, Cu$_{0.5}$Ni$_{0.5}$Rh$_2$S$_4$, and 
CuRh$_2$S$_4$ aligned to the top of the valence band. Note that, despite the general 
similarities, a rigid-band-interpretation is not applicable.}
\label{finedos}
\end{figure}

\begin{figure}
\caption{Calculated total and partial DOS in CuRh$_2$Se$_4$, as shown for CuRh$_2$S$_4$
in Fig. \ref{cus}}
\label{cuse}
\end{figure}

\begin{figure}
\caption{Comparison of calculated XES and partial DOS with experimental spectra of 
CuRh$_2$Se$_4$. The agreement of the main features is within 1 eV (Cu and Se) and even better for Rh.}
\label{cusexes}
\end{figure}

\begin{figure}
\caption{Comparison of XPS VB to Cu L$_3$, and Se M$_{2,3}$ XES in CuRh$_2$Se$_4$.
Note the close alignment of the peaks.}
\label{cusexps}
\end{figure}

\begin{figure}
\caption{Comparison of d-bands from Ni and Cu in CuRh$_2$S$_4$ and NiRh$_2$S$_4$ vs.\ Cu$_{0.5}$Ni$_{0.5}$Rh$_2$S$_4$. 
The significantly different DOS profiles of Ni and Cu d-states in the pure phases 
discounts a rigid-band interpretation. In Cu$_{0.5}$Ni$_{0.5}$Rh$_2$S$_4$ we see that the Cu and Ni d-bands
do not mix very strongly, supporting a ``split-band'' interpretation.}
\label{splitband}
\end{figure}

\onecolumn
%\twocolumn[\hsize\textwidth\columnwidth\hsize\csname @twocolumnfalse\endcsname

\begin{table}
\caption{Transport related quantities and other data}

\begin{tabular}{lcccc}
          & NiRh$_2$S$_4$   & Cu$_{0.5}$Ni$_{0.5}$Rh$_2$S$_4$   & CuRh$_2$S$_4$   & CuRh$_2$Se$_4$\\ \hline
a (\AA)   & 9.71            & 9.75 (assumed)                    & 9.79            & 10.27\\
N(E$_F$) (states/eV-cell) &8.18 &8.43                           & 9.89            & 12.05\\
N(E$_F$) Hagino et al.\cite{hagino} &                           & 12.6            & 13.4\\
v$_F$ ($10^7$ cm/sec) & 2.49 & 2.22                             & 1.79            & 1.75\\
$\hbar\Omega_p$ (eV) & 2.41 & 2.17                              & 1.89            & 1.89\\
T$_c$ (K) & $<$ 2.0  &                                          & 4.70            & 3.483\cite{hagino}\\
\end{tabular}
\label{table1}
\end{table}

%\baselineskip=10pt
%\bibliography{sicl}

\begin{references}
\bibitem{note}  There are several differences between our calculations and those of Ref.\ \onlinecite{matsuno}. 
The most significant difference is that many fewer basis functions
 ($\sim$50/atom) were used in the previous calculation (we used 
$\sim$100 basis functions/atom) and the exchange-correlation potentials were also
different. Furthermore, we used u = 0.385 rather than the ideal position of u = 3/8 for the position
of the S atoms.
\bibitem{foot1}  Rh N$_3$ XES spectra were not measured.
\bibitem{foot2}   Cu L$_3$ and Ni L$_3$ XES are converted to the binding energy scale using our XPS
measurements of corresponding core levels.

labelled
\end{references}
%\bibliographystyle{}
%\newpage 
\clearpage
\epsfxsize=8.0cm\centerline{\epsffile{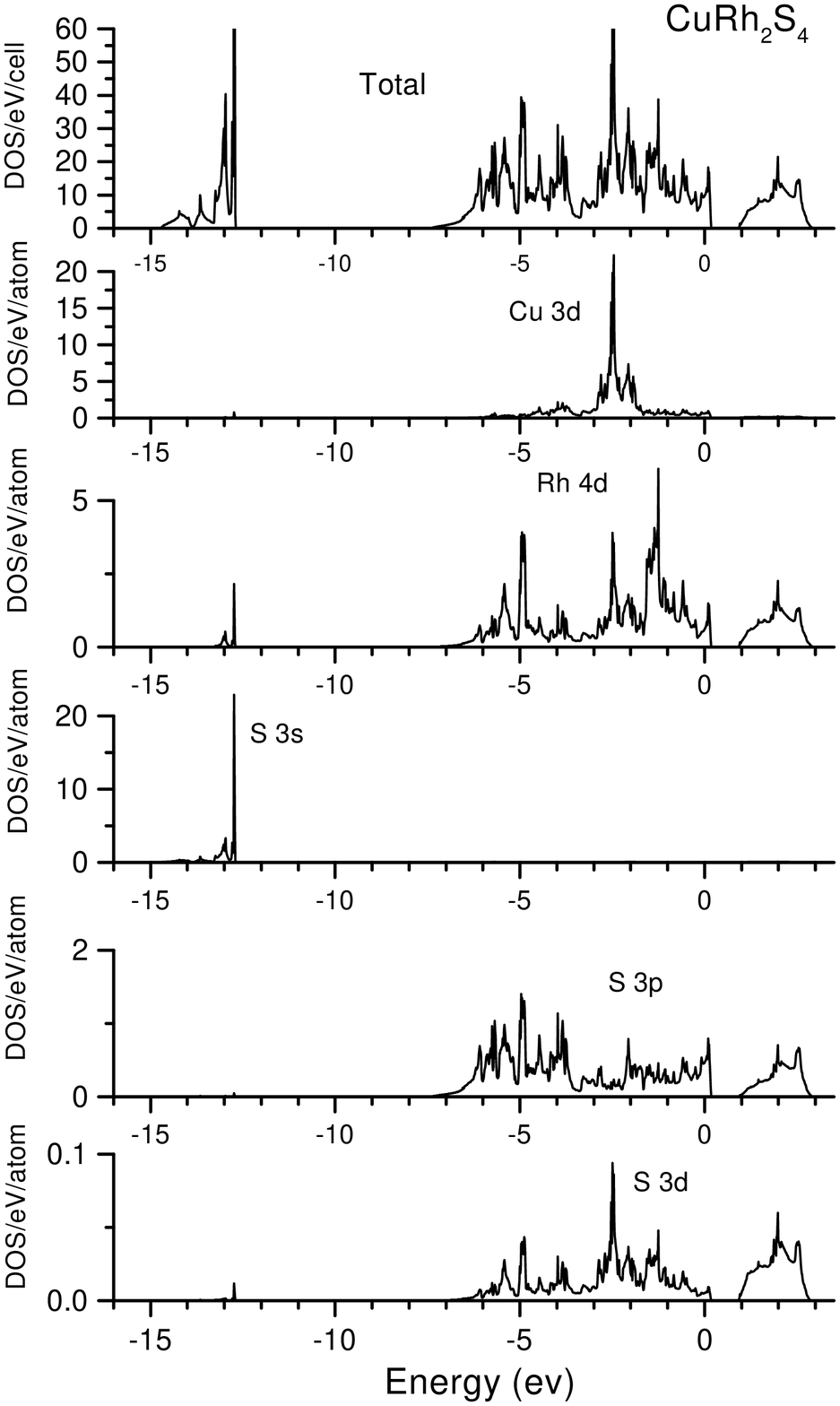}}
\clearpage
\epsfxsize=8.0cm\centerline{\epsffile{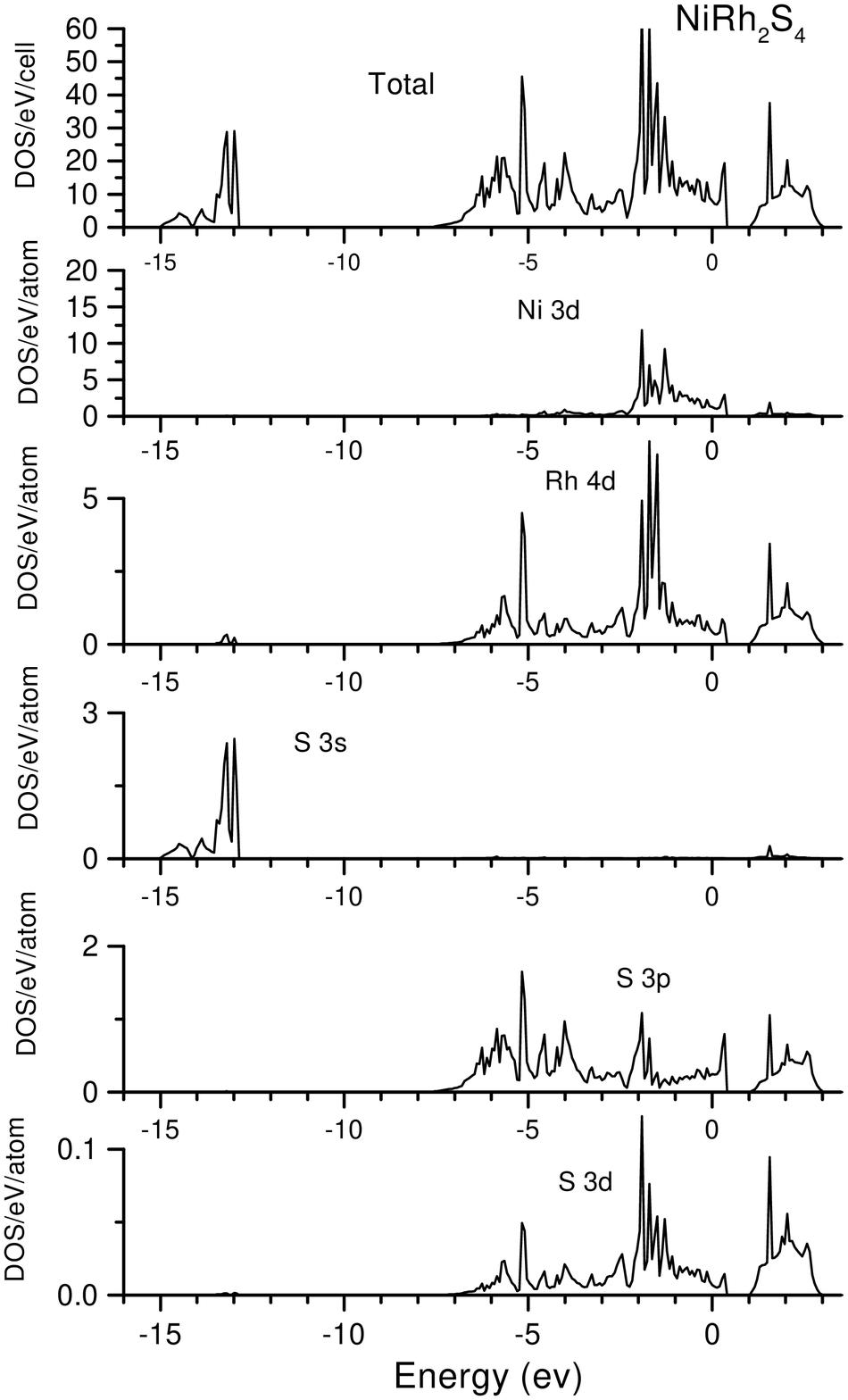}}
\clearpage
\epsfxsize=8.0cm\centerline{\epsffile{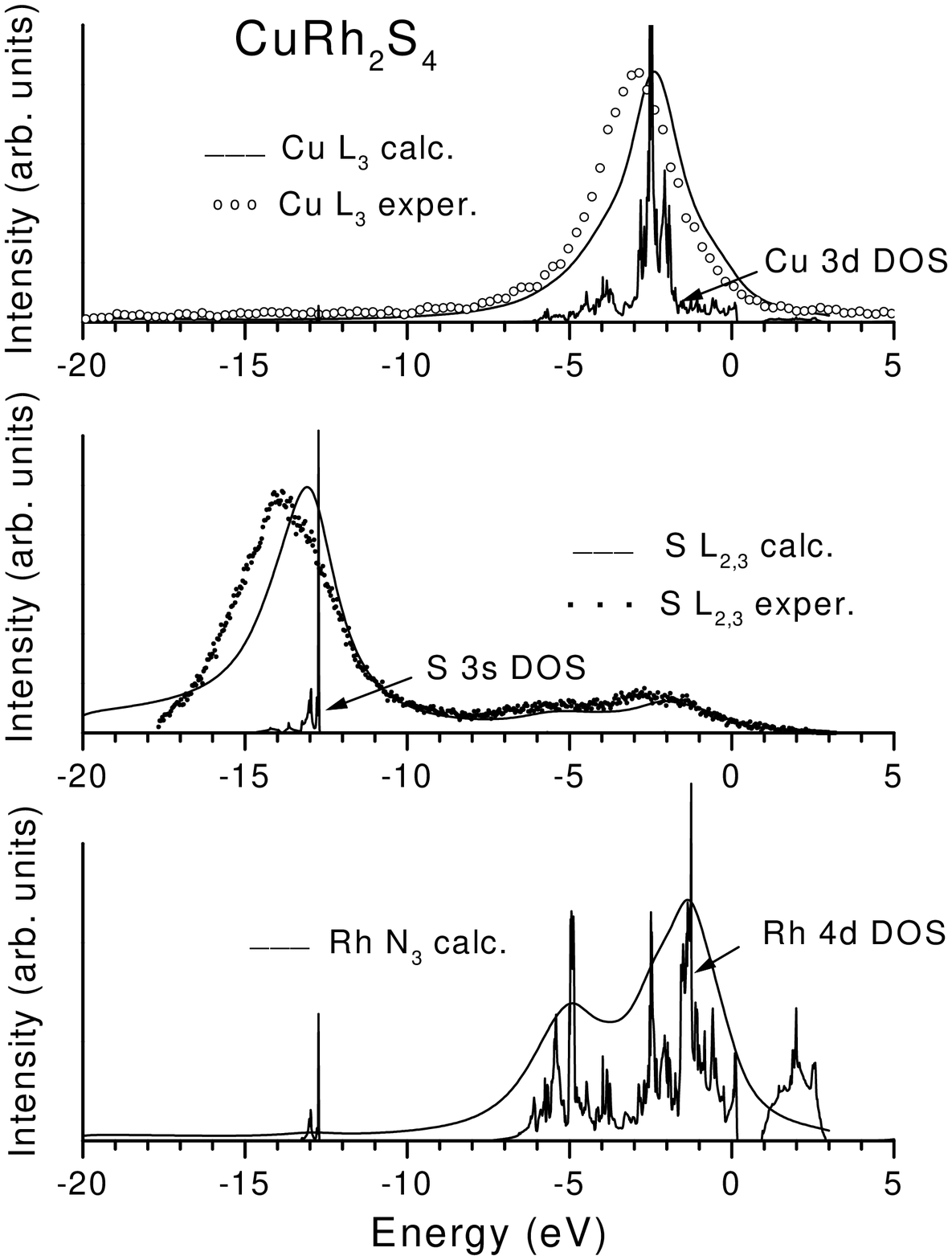}}
\clearpage
\epsfxsize=8.0cm\centerline{\epsffile{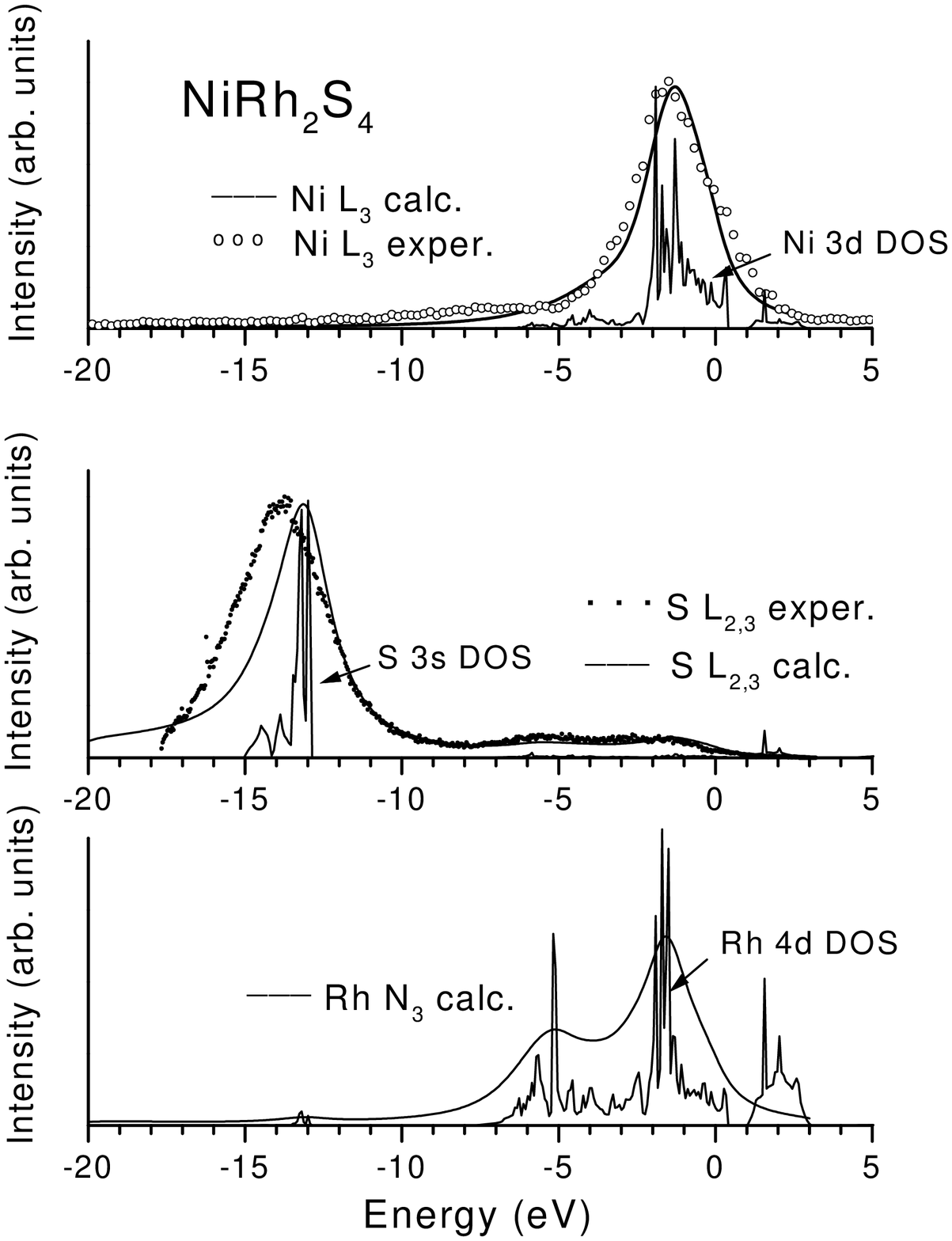}}
\clearpage
\epsfxsize=8.0cm\centerline{\epsffile{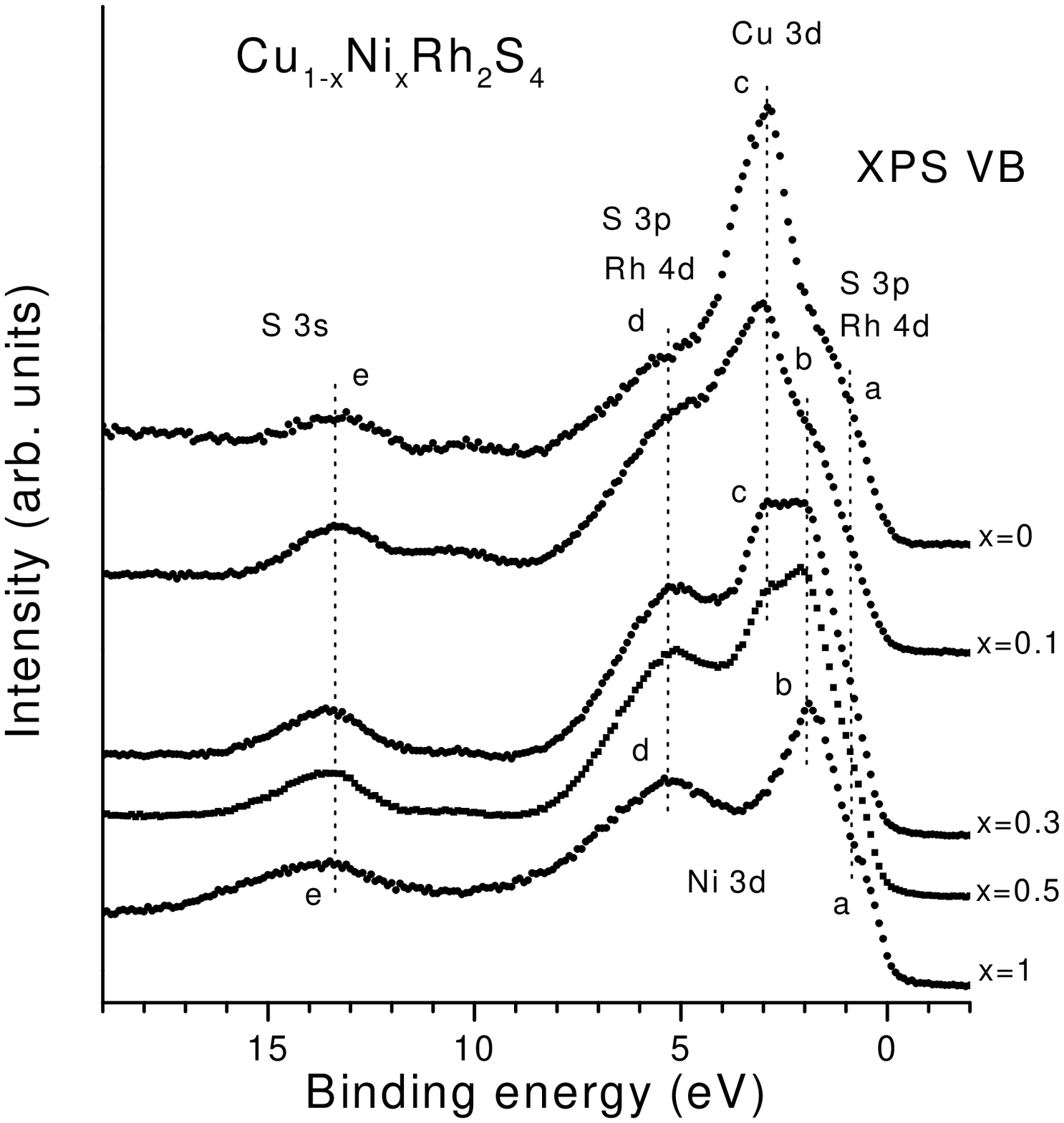}}
\clearpage
\epsfxsize=8.0cm\centerline{\epsffile{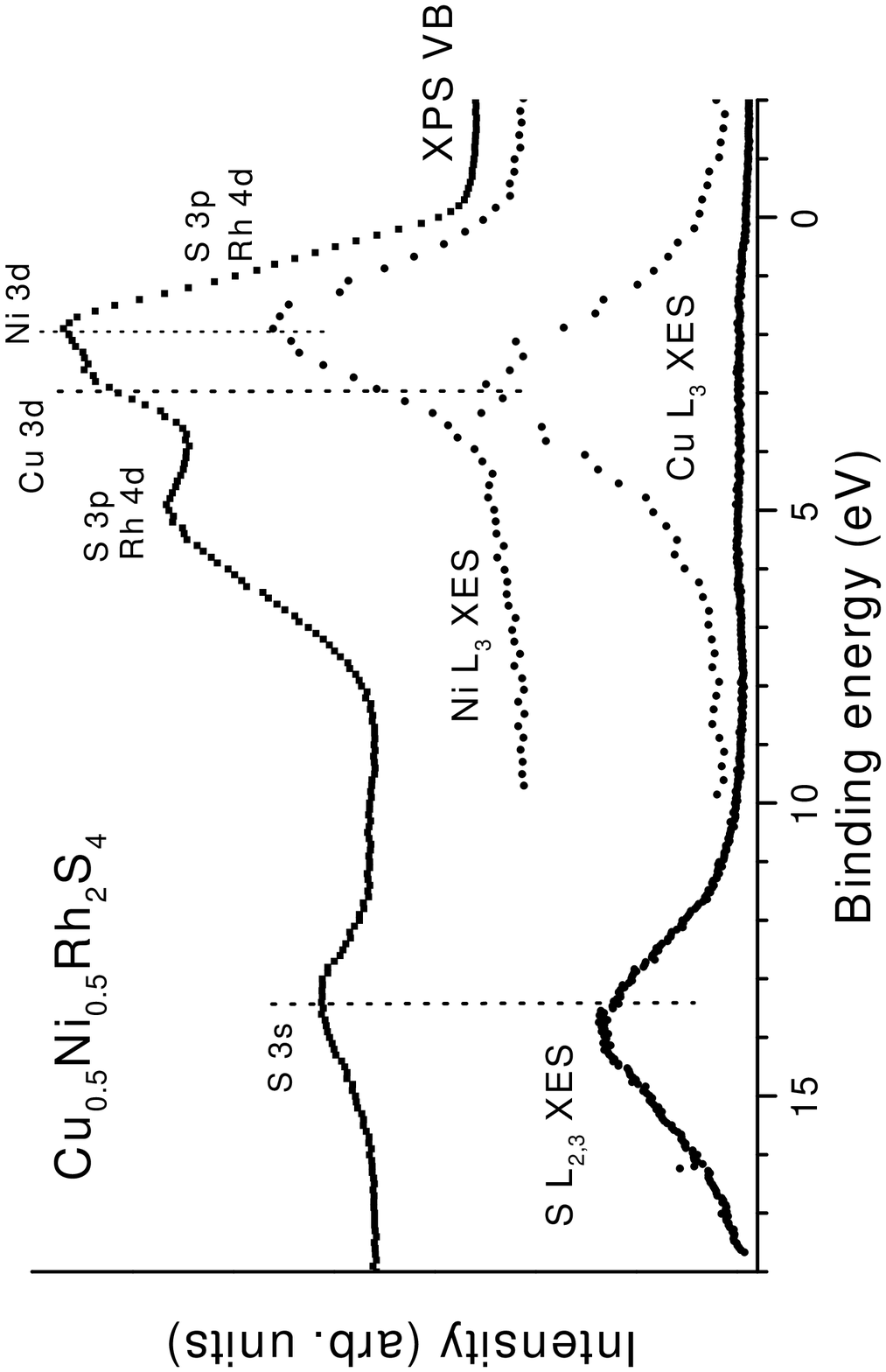}}
\clearpage
\epsfxsize=8.0cm\centerline{\epsffile{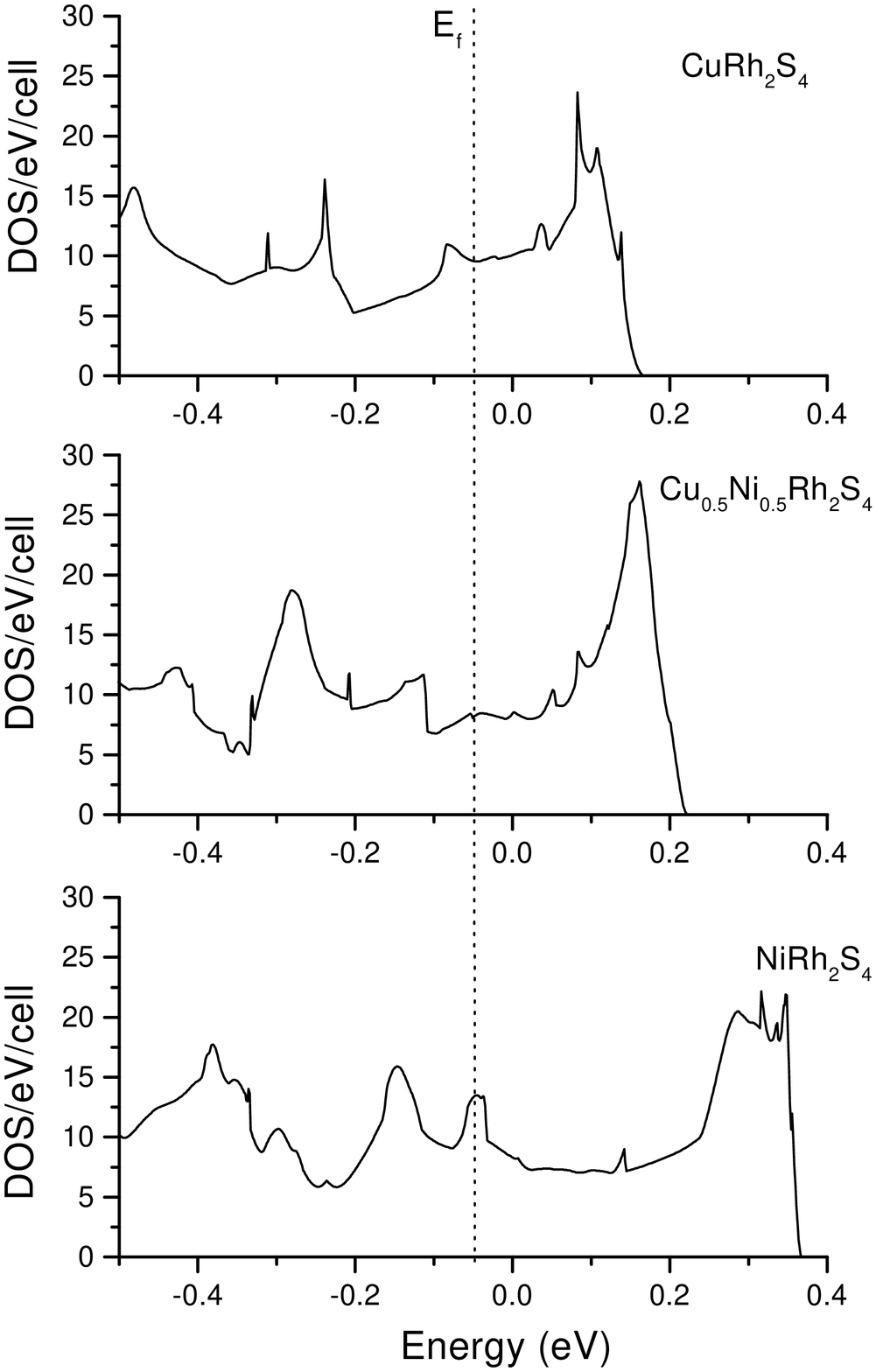}}
\clearpage
\epsfxsize=8.0cm\centerline{\epsffile{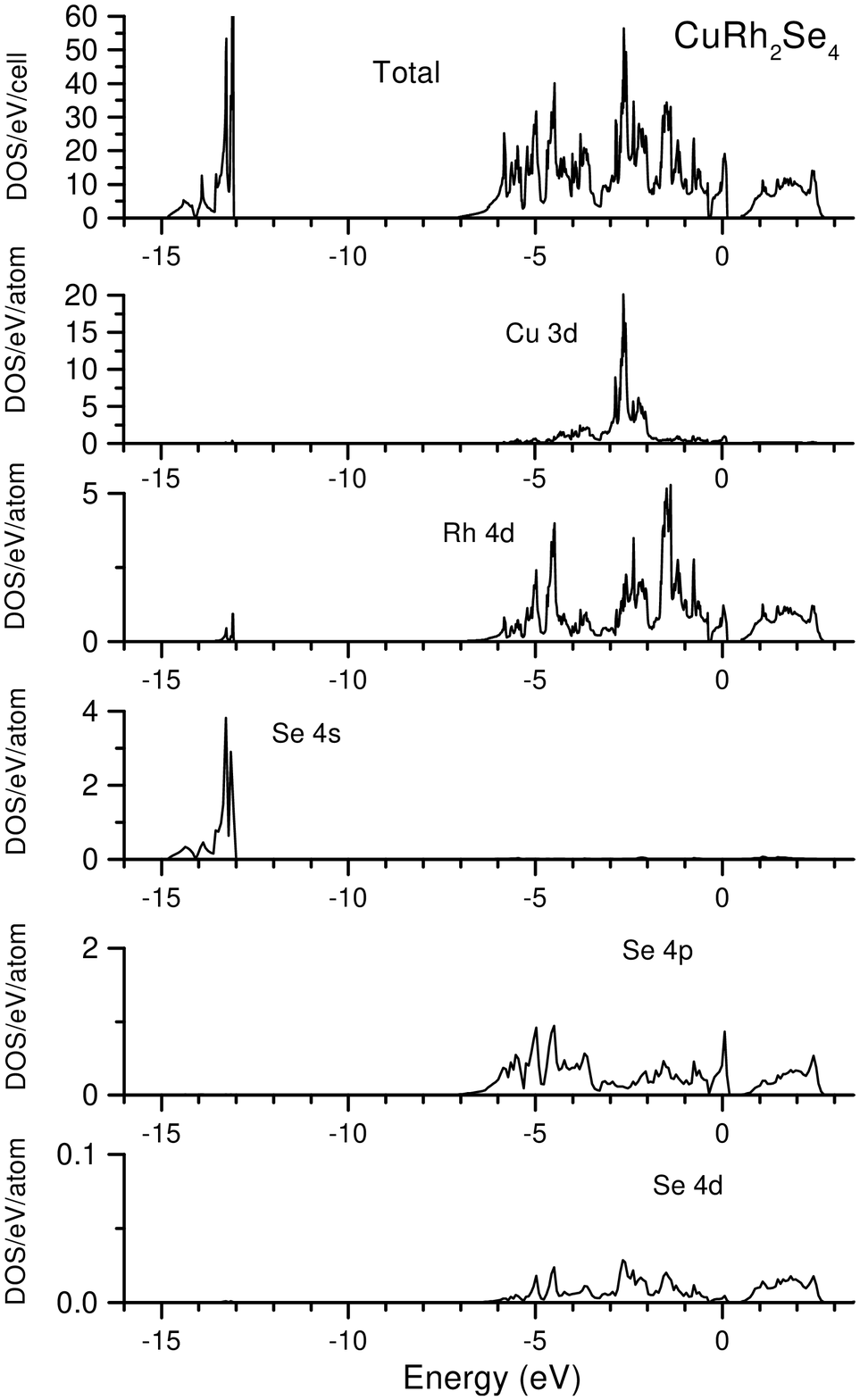}}
\clearpage
\epsfxsize=8.0cm\centerline{\epsffile{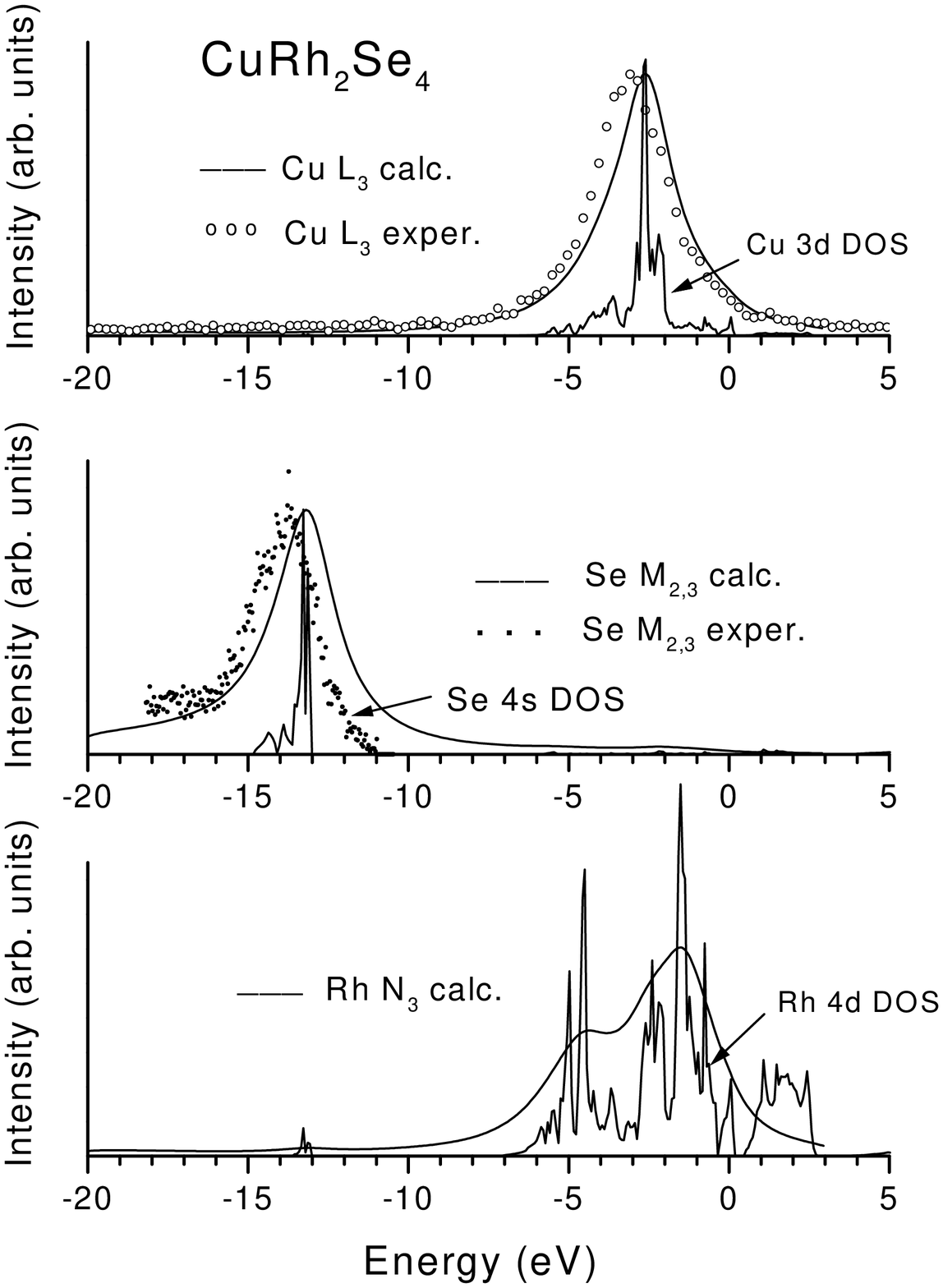}}
\clearpage
\epsfxsize=8.0cm\centerline{\epsffile{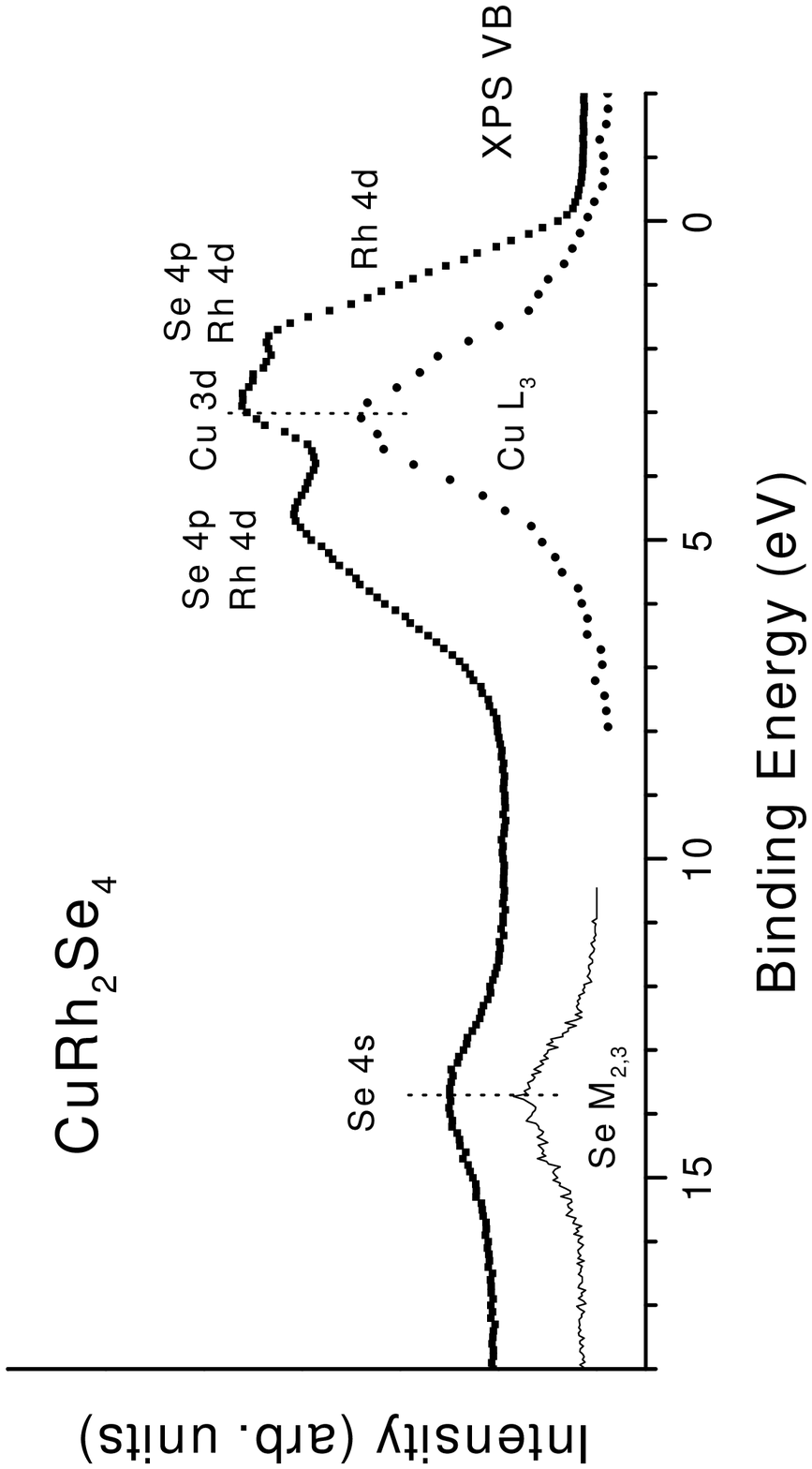}}
\clearpage
\epsfxsize=8.0cm\centerline{\epsffile{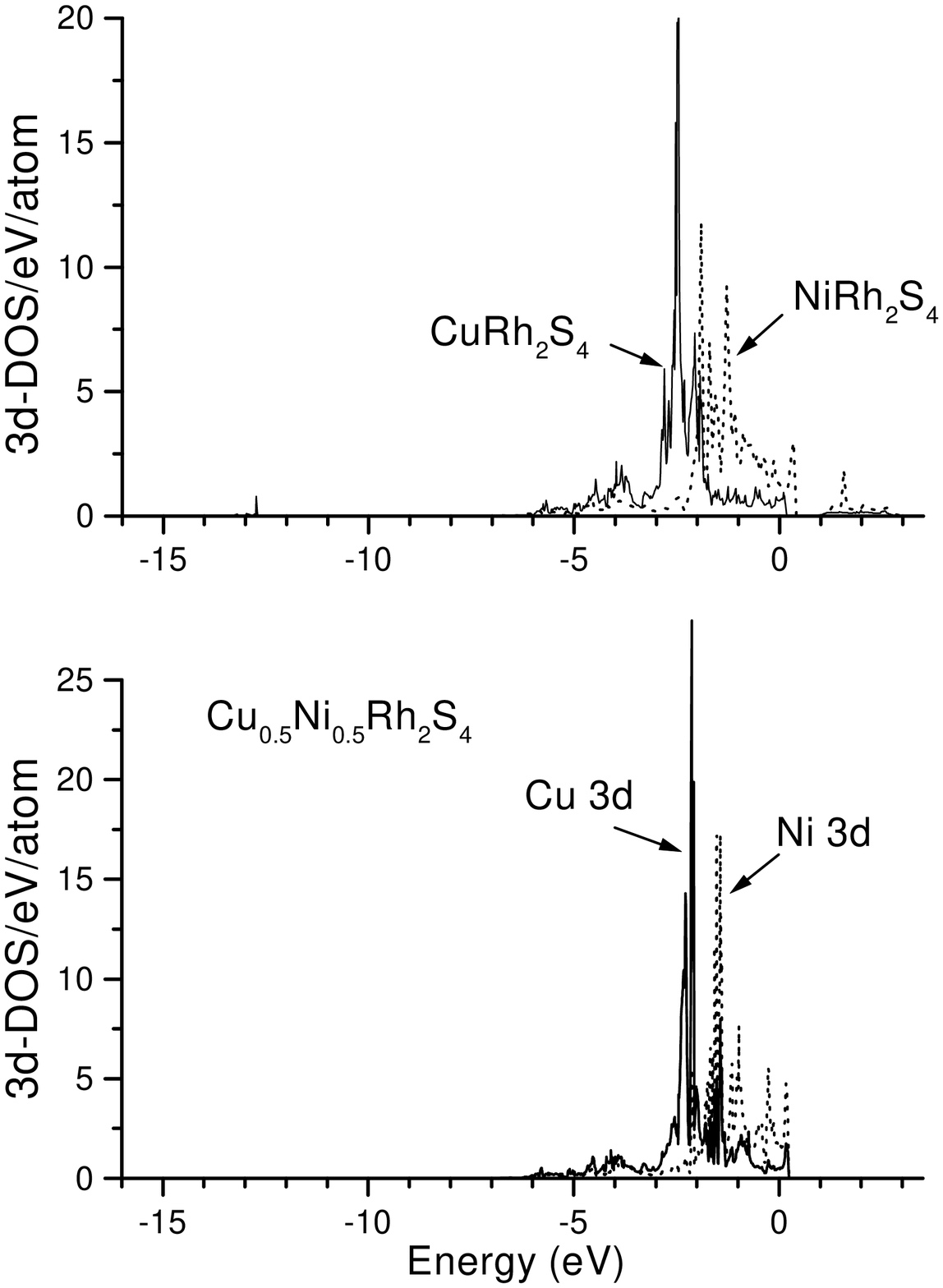}}
\end{document}